# A Comparison of Push and Pull Techniques for AJAX

Engin Bozdag, Ali Mesbah and Arie van Deursen

# A Comparison of Push and Pull Techniques for AJAX

**Engin Bozdag**
Delft University of Technology
The Netherlands
v.e.bozdag@student.tudelft.nl

**Ali Mesbah**
Delft University of Technology
The Netherlands
A.Mesbah@tudelft.nl

**Arie van Deursen**
Delft Univ. of Technology & CWI
The Netherlands
Arie.vanDeursen@tudelft.nl

**Abstract**

AJAX *applications are designed to have high user interactivity and low user-perceived latency. Real-time dynamic web data such as news headlines, stock tickers, and auction updates need to be propagated to the users as soon as possible. However,* AJAX *still suffers from the limitations of the Web's request/response architecture which prevents servers from pushing real-time dynamic web data. Such applications usually use a pull style to obtain the latest updates, where the client actively requests the changes based on a predefined interval. It is possible to overcome this limitation by adopting a push style of interaction where the server broadcasts data when a change occurs on the server side. Both these options have their own trade-offs. This paper explores the fundamental limits of browser-based applications and analyzes push solutions for* AJAX *technology. It also shows the results of an empirical study comparing push and pull.*

## 1. Introduction

The classical style of the web called REST (Representational State Transfer) [5] requires all communication between the browser and the server to be initiated by the client, i.e., the end user clicks on a button or link and thereby requests a new page from the server. In this scheme, each interaction between the client and the server is independent of the other interactions. No 'permanent' connection is established between the client and the server maintains no state information about the clients. This scheme helps scalability, but precludes servers from sending asynchronous notifications.

There are, however, many use cases where it is important to update the client-side interface as soon as possible in response to server-side changes. An auction web site where the users needs to be averted that another bidder has made a higher bid, a stock ticker, a news portal, or a chat-room where new messages are sent immediately to the subscribers, are all examples of such use cases.

Today, such web applications requiring *real-time event notifications* are usually implemented using a pull style, where the client component actively requests the state changes using client-side timeouts. An alternative to this is the push-based style, where the clients subscribe to their topic of interest, and the server publishes the changes to the clients asynchronously every time its state changes.

The recent breed of *Web 2.0* applications dubbed AJAX (Asynchronous JavaScript and XML) [7] is designed to have high user interactivity and low user-perceived latency [13]. Introducing the push style into AJAX systems [10] can further improve the responsiveness of such applications towards end users.

However, implementing such push solution for web applications is not trivial, mainly due to the limitations of the HTTP protocol. This research explores the fundamental limits of browser-based applications in providing real-time data. We explore how real-time event notification can be added to today's AJAX technology and compare the pull and push approaches by conducting an empirical study to find out the actual trade-offs of each approach.

This paper is further organized as follows. Section 2 shows current techniques to implement HTTP based push and discusses the BAYEUX protocol [17], which tries to bring a standard to HTTP push. Section 3 explains our setup for the push-pull experiment. Section 4 presents the results of the empirical study involving push and pull. Section 5 discusses the results of the study. Section 6 summarizes related work on this area. Finally, Section 7 ends this paper with concluding remarks.

## 2. Web-based Real-time Event Notification

### 2.1. AJAX

AJAX [7] is an approach to web application development utilizing a combination of established web technologies: standards-based presentation using XHTML and CSS, dynamic display and interaction using the Document Object Model, data interchange and manipulation, asynchronous data retrieval using XMLHttpRequest, and JavaScript binding everything together. XMLHttpRequest is an API implemented by most modern web browser scripting engines to transfer data to and from a web server using HTTP, by estab-lishing an independent communication channel in the background between a web client and server.

It is the combination of these technologies that enables us to adopt principal software engineering paradigms, such as component- and event-based, for web application development. Our earlier work [13] on an architectural style for AJAX, called SPIAR, gives an overview of the new way web applications can be architected using AJAX. Adopting AJAX has become a serious option not only for newly developed applications, but also for migrating [14] existing web sites to increase the responsiveness. The evolution of web and the advent of *Web 2.0*, and AJAX in particular, is making the users experience similar to using a desktop application. Well known examples include Gmail, and the new version of Yahoo! Mail.

The REST style makes a server-initiated HTTP request impossible. Every request has to be initiated by a client, precluding servers from sending asynchronous notifications without a request from the client [11]. There are several solutions used in the practice that still allow the client to receive (near) real-time updates from the server. In this section we will analyze some of such solutions.

## 2.2. HTTP Pull

Most AJAX applications check with the server at regular user-definable intervals known as *Time to Refresh* (TTR). This check occurs blindly regardless of whether the state of the applications has changed.

In order to achieve high data accuracy and data freshness, the pulling frequency has to be high. This, in turn, induces high network traffic and possibly unnecessary messages. The application also wastes some time querying for the completion of the event, thereby directly impacting the responsiveness to the user. Ideally, the pulling interval should be equal to the *Publish Rate* (PR), i.e., rate at which the state changes. If the frequency is too low, the client can miss some updates.

This scheme is frequently used in web systems, since it is robust, simple to implement, allows for offline operation, and scales well to high number of subscribers [8]. Mechanisms such as Adaptive TTR [3] allow the server to change the TTR, so that the client can pull on different frequencies, depending on the change rate of the data. This dynamic TTR approach in turn provides better results than a static TTR model [18]. However, it will never reach complete data accuracy, and it will create unnecessary traffic.

## 2.3. HTTP Streaming

HTTP Streaming is a basic and old method that was introduced on the web first in 1992 by Netscape, under the name 'dynamic document' [15]. HTTP Streaming comes in two forms namely, Page and Service Streaming.

**Page Streaming**
This method simply consists of streaming server data in the response of a long-lived HTTP connection. Most web servers do some processing, send back a response, and immediately exit. But in this pattern, the connection is kept open by running a long loop. The server script uses event registration or some other technique to detect any state changes. As soon as a state change occurs, it streams the new data and flushes it, but does not actually close the connection. Meanwhile, the browser must ensure the user-interface reflects the new data, while still waiting for response from the server to finish.

**Service Streaming**
Service Streaming relies on the XMLHttpRequest object. This time, it is an XMLHttpRequest connection that is long-lived in the background, instead of the initial page load. This brings some flexibility regarding the length and frequency of connections. The page will be loaded normally (one time), and streaming can be performed with a predefined lifetime for connection. The server will loop indefinitely just like in page streaming, the browser has to read the latest response (*responseText*) to update its content.

## 2.4. COMET and the BAYEUX Protocol

The application of the Service Streaming scheme under AJAX is now known as Reverse AJAX or COMET [16]. COMET enables the server to send a message to the client when the event occurs, without the client having to explicitly request.

As a response to the lack of communication standards [13] for AJAX applications, the Cometd group[1] released a COMET protocol draft called BAYEUX [17]. The BAYEUX message format is defined in JSON (JavaScript Object Notation)[2] which is a data-interchange format based on a subset of the JavaScript Programming Language. The protocol has recently been implemented and included in a number of web servers including Jetty[3] and IBM Websphere. This protocol currently provides a connection type called *Long Polling* for HTTP push, which is implemented in Jetty's Cometd library[4].

**Long Polling** (also known as Asynchronous-Polling) is a mixture of pure server push and client pull. After a subscription to a channel, the connection between the client and the server is kept open, for a defined period of time (by default 45 seconds). If no event occurs on the server side, a timeout occurs and the server asks the client to reconnect asynchronously. If an event occurs, the server sends the data to the client and the client reconnects.

This protocol follows the 'topic-based' [4] publish-subscribe scheme, which groups events according to their topic (name) and map individual topics to distinct communication channels. Participants subscribe to individual topics, which are identified by keywords. Like many modern topic-based engines, BAYEUX offers a form of *hierarchical addressing*, which permits programmers to organize topics according to containment relationships. It also allows topic names to contain *wildcards*, which offers the possibility to subscribe and publish to several topics whose names match a given set of keywords. BAYEUX defines the following phases in order to establish a COMET connection:

1. Client performs a handshake with the server, receives a client id and list of supported connection types (IFrame, long-polling, etc.).
2. Client sends a connection request with its id and its preferred connection type.
3. Client later subscribes to a channel and receives updates

---

[1] http://www.cometd.com
[2] http://www.json.org
[3] http://www.mortbay.org
[4] http://www.mortbay.org

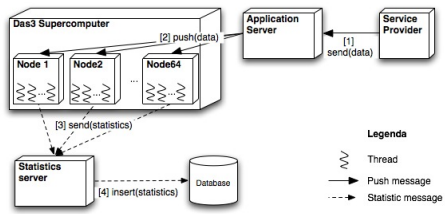

**Figure 1. Experimental Environment**

In the remainder of this paper, we will use BAYEUX as the protocol for server push, and compare its performance with a pure pull based solution.

## 3. Experimental Design

In this section we will present our experimental setup.

### 3.1. Goals and Setup

The goals of our experiment consist of exploring the actual performance trade-offs of a COMET push implementation and compare it to a pure pull approach on the web by conducting a controlled empirical study. The experiment has to be repeatable for push and pull but also for different input variables such as number of users, number of published messages and intervals.

We aim at achieving these goals by:

- creating a push application consisting of the client and the server parts,
- creating the same application for pull,
- creating an application which publishes a variable number of data items at certain intervals,
- mimicking many concurrent web clients operating on each application,

- gathering data and measuring: the mean time it takes for clients to receive a new published message, the load on the server, number of messages sent or retrieved, the effects of changing the data publish rate and number of users,
- analyzing and explaining the measurements found.

To see how the application server reacts to different conditions, we use different combinations of three variables:

- Number of concurrent users (100, 200, 350, 500, and 1000). The variation helps to find a maximum number of users the server can handle simultaneously and 1000 seemed to be the upper-bound for our test. This is because the server was already running on 100% CPU with 1000 users. We also tried 2000 and 5000 users, however the server was so saturated that it was not able to send any updates anymore.
- Publish interval (5, 10, 15, and 50 seconds): The frequency of the publishing updates is also important. Because of the *long polling* implementation in BAYEUX (See Section 2), the system should act more like pure pull when the publish interval is small, and more like pure push when it is bigger. We chose the interval 50 seconds, because the client timeout of BAYEUX protocol is 45 seconds, and we expect this interval to cause many disconnects, hence affecting the performance.
- Push or Pull: We also made an option in our test script that allowed us to switch between pull and push. To make the total number of combinations smaller, we set the pull interval as 15 seconds.
- Total number of messages: For each combination, we generated a total of 10 publish messages.

### 3.2. Tools

In order to simulate a high number of clients, we evaluated several open source solutions. *Grinder*[5] seemed to be a good option, providing an internal TCPProxy, allowing to record events sent by the browser and later replay them. It also provided scripting support, which allowed us to create a script that simulates a browser connecting to the push server, subscribing to a particular stock channel and receiving push data continuously. In addition, Grinder has a built-in feature that allows us to create multiple threads of a simulating script.

Because of the distributed nature of the simulated clients on different nodes, we used Log4J's SocketServer[6] to set up a logging server that listens for incoming log messages. The clients then send the log messages using the `SocketAppender`.

We used *TCPDump*[7] to record the number of TCP (HTTP) packets sent to and from the server. We also cre-ated a script that uses the UNIX *top* utility[8] to record the CPU usage of the application server. This was necessary to observe the scalability and performance of each approach.

---
[5] http://grinder.sourceforge.net
[6] http://logging.apache.org/log4j/docs/
[7] http://www.tcpdump.org/

### 3.3. Sample Application

In order to respond to publish events and create client-side processing, we developed a Stock Ticker application.

**The Push version** consists of a JSP page which uses Dojo's Cometd library[9] to subscribe to a channel and receive the Stock data. We use Rico[10] to give color effects to different data values on the web interface. For the server side, we developed a Java Servlet (PushServlet) that pushes the data into the browsers using the Cometd library. The PushServlet manages the client connections, receives data from back-end, and publishes it to the clients.

**The pull version** has also one JSP page, but instead of Cometd, it uses the normal *bind* method of Dojo to request data from the server. The pull nature was set using the standard `setInterval` JavaScript method. On the server, a PullServlet was made which keeps an internal stock object (the most recent one) and simply handles every incoming request the usual way.

**The Service Provider** Java application was created which uses the HTTPClient library[11] to publish stock data to the Servlets. The number of publish messages as well as the interval at which the messages are published are configurable.

**Simulating clients** To simulate many concurrent clients we use the TCPProxy to record the actions of the JSP/Dojo client pages for push and pull and create scripts for each in Jython[12]. Jython is an implementation of the high-level, dynamic, object-oriented language Python, integrated with the Java platform. It allows the usage of Java objects in a Python script and is used by Grinder to simulate web users. In our tests, Jython scripts are actually imitating the JSP/Dojo client pages.

### 3.4. Testing Environment

We use the Distributed ASCI Supercomputer 3 (DAS3)[13] to run various numbers of web clients on different distributed nodes. The DAS3 cluster at the Delft University consists of 68 dual-CPU 2.4 GHz AMD Opteron DP 250 compute nodes, each having 4 GB of memory. The cluster is equipped with 1 and 10 Gigabit/s Ethernet, and runs Scientific Linux 4.

The application server runs on a Pentium IV, 3 Ghz (Hyperthreading) machine with 1 Gb memory, and Linux Fedora as its Operating System. We use Jetty 6.1.2 as our application server, because it is the only open-source Java EE application server that currently implements the COMET BAYEUX protocol. Jetty uses Java's new IO package (NIO). NIO package follows the event-driven design, which allows the processing of each task as a finite state machine (FSM). As the number of tasks reach a certain limit, the excess tasks are absorbed in the server's event queue. The throughput remains constant and the latency shows a linear increase. The Event-driven design is supposed to perform significantly better than thread-concurrency model [20, 21].

The connectivity between the server and DAS3 nodes is through a 100 Mbps ethernet connection.

### 3.5. Sequence of events

A routine test run consists of the following steps (See Figure 1):

1. The Service Provider publishes the stock data to the application server via an HTTP POST request, in which the creation date, the stock item id, and the stock data are specified.
2. For push: The application server *pushes* the data to all the subscribers of that particular stock. For pull: the application server updates the internal stock object, so that when clients send pull requests, they get the latest data.
3. Each simulated client logs the responses (after some calculation) and sends it to the statistics server. Grinder also processes the data from each client and sends the statistics, such as response time, to the statistics server, which runs on a separate machine.

It is worth noting that we use a combination of the 64 DAS3 nodes and Grinder threads to simulate different numbers of users.

### 3.6. Data Analysis

We created a Data Analyzer that reads the data from Grinder and Logging Server logs and writes all the info into a database using Hibernate[14]. This way, different views of the data can be obtained easily using queries to the database.

## 4. Results

In the following subsections, we present the results which we obtained using the combination of variables mentioned in 3.1. Figures 2–5 depict the results. Note that for each number of clients on the x-axis, the five publish intervals in seconds (5, 10, 15, 20, 50) are presented.

---

[8] http://www.unixtop.org/
[9] http://dojotoolkit.org/
[10] http://www.openrico.org/
[11] http://jakarta.apache.org/commons/httpclient/
[12] http://www.jython.org
[13] http://www.cs.vu.nl/das3/overview.shtml
[14] http://www.hibernate.org

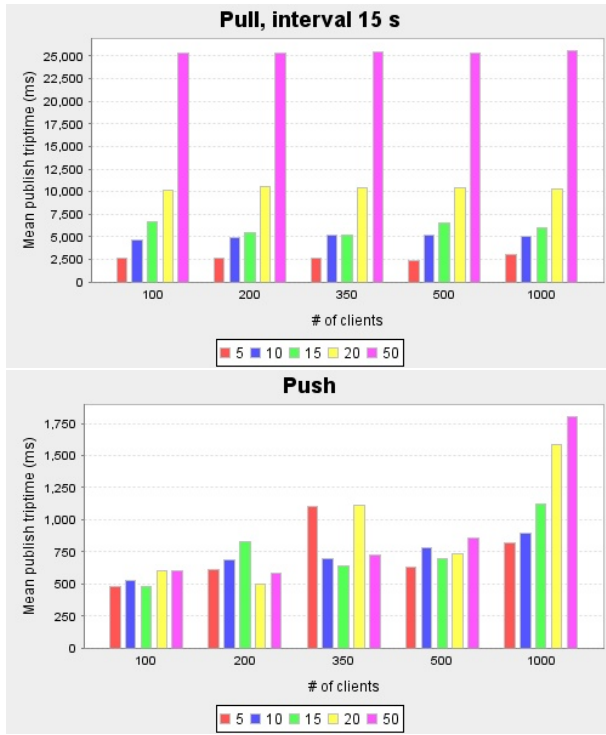

**Figure 2. Mean publish triptime.**

## 4.1. Publish triptime

We define triptime as follows:

*Triptime = | Data Creation Date − Data Receipt Date |*

*Data Creation Date* is the date on the Service Provider (Publisher) the moment it creates a message, and *Data Receipt Date* is the date on the client the moment it receives the message. Triptime shows how long it takes for a publish message to reach the client and can be used to find out how fast the client gets notified with the latest events. Note that it is very important to synchronize the datetime for both the Service Provider and the clients.

Figure 2 shows the mean publish triptime versus the total number of clients, for both pull and push techniques.

## 4.2. Server Performance

Since push is stateful, we expect it to have some administration costs on the server side, using resources. In order to compare this with pull, we measured the CPU usage for both approaches. Figure 3 shows the mean server CPU usage as the number of clients grow, for push and pull.

## 4.3. Received Publish Messages

To see how pull compares to pure push in message overhead, we published a total of 10 messages and we counted the total number of (non unique) messages received on the client side. Figure 4 shows the mean number of received non-unique publish items versus the total number of clients, for both push and pull. Note that if a pull client makes a request while there is no new data, it will receive the same item multiple times. This way a client might receive more than 10 messages.

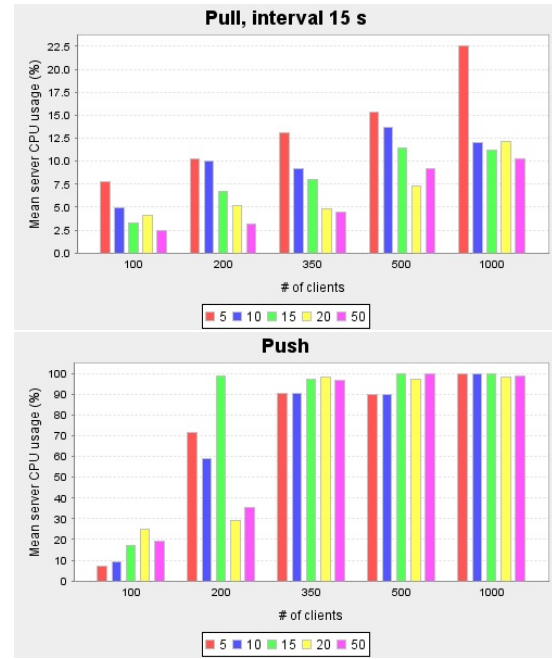

**Figure 3. Server application CPU usage.**

## 4.4. Received Unique Publish Messages

It is also interesting to see how many of the 10 messages we have published reach the clients. This way we can determine if the clients miss any publish items. Figure 5 shows the mean number of received unique publish items versus total number of clients.

## 5. Discussion

### 5.1. Data Coherence

We define a piece of data as coherent, if the data on the server and the client is synchronized. We check the data coherence of both approaches by measuring the triptime. As we can see in Figure 2, the triptime is, at most, 1,750 milliseconds with push. In Pull, this can go up to 25 seconds. This shows us that pull is not as responsive as push, and if we need high data coherence, we should always choose the push approach. In Figure 2 we also see that with 1000 users and a publish interval of 50 seconds, the triptime increases noticeably. With such a big interval, no response is being sent to the client, and the client is waiting for data, thus occupying a thread. This makes it hard for other clients to reconnect and get new data, which increases the triptime. With an interval of 5 seconds, the triptime is lower. This is because the clients are quickly receiving responses and disconnecting. This makes some threads available, which makes it possible for other clients to connect.

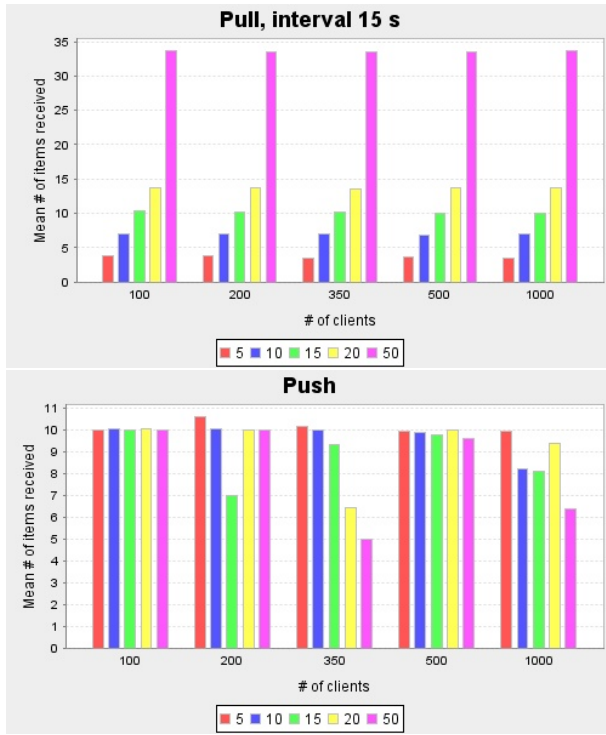

**Figure 4. Mean Number of Received Publish Items.**

## 5.2. Server Performance

One of the main issues of all distributed systems and in particular that of web-based applications is scalability and performance. As it is depicted in Figure 3, the pull style has a much better performance compared to push and this is valid even for small number of users (e.g., 100). With push, when the number of clients is increased to 350, the server is practically saturated, i.e., CPU is running at almost 100%. This is mainly due to the fact that the push server has to maintain all the state information about the clients and also manage the corresponding threads and connections. A push server based on long polling also needs to generate numerous request/response cycles to keep the connection alive, which impact the resources. With pull only the publish interval has a direct measurable effect on the performance. This shows us that if we want to use a push implementation even for a couple of hundreds of users, some load balancing solution and multiple servers are needed.

## 5.3. Network Performance

As we mentioned in Section 2.2, in a pure pull system, the pulling frequency has to be high to achieve high data accuracy and data freshness. If the frequency is higher than the data generation interval, the pull client will pull the same data more than once, leading to some overhead.

In Figure 4 we see that with a publish interval of 50 seconds, pull clients receive approximately 35 messages, while we published only 10. In the same figure we see that Push clients received approximately a maximum of 10 messages. This means that, more than 2/3 of total number of pull requests were unnecessary. Furthermore, we see that the number of packages received does not depend on the number of clients.

If we look at Push graph in Figure 5, we notice that as the number of users increase, not all clients receive all 10 messages. The number of correctly received messages is quite well with 100 users, but, unlike the pure pull approach, it begins to degrade as the users increase. This shows that Jetty's Cometd implementation is not stable and scalable enough.

## 5.4. Data Misses

According to Figure 5, if the publish interval is 20 or 50 (i.e., larger than the pull interval), the client receives all the messages. However as we have discussed in the previous subsection, this will generate an unnecessary number of messages. Looking at the figure again, we see that when the pull interval is smaller than the publish interval, the clients will miss some updates, regardless of the number of users. So, with the pull approach, we need to know the *exact* publish interval. However the publish interval tends to change, which makes it difficult for a pure pull implementation. With push, when the number of clients is small, the client will receive all the messages. However if the number of clients increases, and the publish interval is large, some data loss starts to occur. This is again due to high number of idle threads, which affects the server performance.

## 5.5. Threats to Validity

We use several tools to obtain the data. The shortcomings and the problems of the tools themselves can have an effect on the outcome. In addition, implementation issues in the application server Jetty 6.1.2 might lead to the high CPU usage.

Another threat is the pull interval. We use only 1 pull interval, namely 15 seconds. Different pull intervals might have an influence on the performance of the server and the data coherence.

Clients can also have different environments (i.e.: the browser they use, the bandwidth they have, etc.). This can have an influence on the triptime variable. In order to avoid

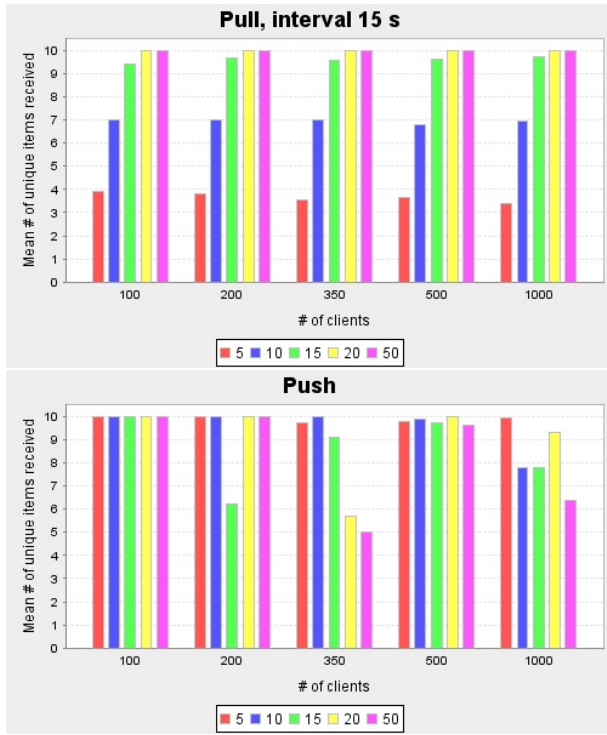

**Figure 5. Mean Number of Received Unique Publish Items.**

that, we used the same test-script in all the simulated clients and allocated the same bandwidth.

The time can also be a threat to validity. To measure the trip-time, the difference between the data creation date and data receipt date is calculated. However if the time on the server and the clients are different, this might give a false trip time. In order to prevent this problem, we made sure that the time on server and client machines are synchronized by using the same time server.

We measure the data coherence by taking the trip time. However, the data itself must be 'correct', i.e., the received data must be the same data that has been sent by the server. We rely on HTTP in order to achieve this "data correctness". However, additional experiments must include a self check to ensure this requirement.

## 6. Related Work

There are a number of papers that discuss server-initiated events, known as *push*, however, most of them focus on client/server distributed systems and non HTTP multimedia streaming or multi-casting with a single publisher [1, 9, 6, 2, 19]. The only work that focuses on AJAX is the white-paper of Khare [10]. Khare discusses the limits of the pull approach for certain AJAX applications and mentions several use cases where a push application is much more suited.

However, the white-paper does not mention possible issues with this *push* approach such as scalability and performance. Khare and Taylor [11] propose a push approach called AR-RESTED. Their asynchronous extension of REST, called A+REST, allows the server to broadcast notifications of its state changes. The authors note that this is a significant implementation challenge across the public Internet.

The research of Acharya *et al.* [1] focuses on finding a balance between push and pull by investigating techniques that can enhance the performance and scalability of the system. According to the research, if the server is lightly loaded, pull seems to be the best strategy. In this case, all requests get queued and are serviced much faster than the average latency of publishing. The study is not focused on HTTP.

Bhide *et al.* [3] also try to find a balance between push and pull, and present two dynamic adaptive algorithms: *Push and Pull* (PaP), and *Push or Pull* (PoP). According to their results, both algorithms perform better than pure pull or push approaches. Even though they use HTTP as messaging protocol, they use custom proxies, clients, and servers. They do not address the limitations of browsers nor do they perform load testing with high number of users.

Hauswirth and Jazayeri [8] introduce a component and communication model for push systems. They identify components used in most *Publish/Subscribe* implementations. The paper mentions possible problems with scalability, and emphasizes the necessity of a specialized, distributed, broadcasting infrastructure.

Eugster *et al.* [4] compare many variants of *Publish/Subscribe* schemes. They identify three alternatives: *topic-based*, *content-based*, and *type-based*. The paper also mentions several implementation issues, such as events, transmission media and qualities of service, but again the main focus is not on web-based applications.

Flatin [12] compares push and pull from the perspective of network management. The paper mentions the publish/subscribe paradigm and how it can be used to conserve network bandwidth as well as CPU time on the management station.suggests the 'dynamic document' solution of Netscape [15], but also a 'position swapping' approach in which each party can both act as a client and a server. This solution, however, is not applicable to web browsers. Making a browser act like a server is not trivial and it induces security issues.

As far as we know, there has been no empirical study conducted to find out the actual trade-offs of applying pull/push on browser-based or AJAX applications.

## 7. Conclusion

In this paper we have compared pull and push solutions for achieving web-based real time event notification. The contributions of this paper include the experimental design, a reusable implementation of a sample application in push and pull style as well as a measurement framework, and the experimental results.

Our experiment shows that if we want high data coherence and high network performance, we should choose the push approach. However, push brings some scalability issues; the server application CPU usage is 7 times higher as in pull. According to our results, the server starts to saturate at 350-500 users. For larger number of users, load balancing and server clustering techniques are unavoidable.

With the pull approach, achieving total data coherence with high network performance is very difficult. If the pull interval is higher than the publish interval, some data miss will occur. If it is lower, network performance will suffer. Pull performs well only if the pull interval equals to publish interval. However, in order to achieve that, we need to know the exact publish interval beforehand. However, the publish interval is rarely static and predictable. This makes pull useful only in situations where the data is published frequently according to some pattern.

These results allow engineers to make rational decisions concerning key parameters such as pull and push intervals, in relation to, e.g., the anticipated number of clients. Furthermore, the experimental design allows them to repeat similar measurements for their own (existing or to be developed) applications.

Our future work includes adopting a hybrid approach that combines pull and push techniques for AJAX applications to gain the benefits of both approaches. We also intend to extend our testing experiments to different versions of Jetty and alternative push server implementations, for example ones that are based on holding a permanent connection (e.g., Lightstreamer[15]) as opposed to the long polling approach discussed in this paper. Additional experiments with a variety of pull intervals are also desired.

**Acknowledgments** Partial support was received from SenterNovem, project Single Page Computer Interaction (SPCI), in collaboration with Backbase.

# References

[1] S. Acharya, M. Franklin, and S. Zdonik. Balancing push and pull for data broadcast. In *SIGMOD '97: Proceedings of the 1997 ACM SIGMOD international conference on Management of data*, pages 183–194. ACM Press, 1997.

[2] M. Ammar, K. Almeroth, R. Clark, and Z. Fei. Multicast delivery of web pages or how to make web servers pushy. Workshop on Internet Server Performance, 1998.

[3] M. Bhide, P. Deolasee, A. Katkar, A. Panchbudhe, K. Ramamritham, and P. Shenoy. Adaptive push-pull: Disseminating dynamic web data. *IEEE Trans. Comput.*, 51(6):652–668, 2002.

[4] P. T. Eugster, P. A. Felber, R. Guerraoui, and A.-M. Kermarrec. The many faces of publish/subscribe. *ACM Comput. Surv.*, 35(2):114–131, 2003.

[5] R. T. Fielding and R. N. Taylor. Principled design of the modern web architecture. *ACM Trans. Inter. Tech.*, 2(2):115–150, 2002.

[6] M. Franklin and S. Zdonik. data in your face: push technology in perspective. In *SIGMOD '98: Proceedings of the 1998 ACM SIGMOD international conference on Management of data*, pages 516–519. ACM Press, 1998.

[7] J. Garrett. Ajax: A new approach to web applications. Adaptive Path: http://adaptivepath.com/publications/essays/archives/000385.php, 2005.

[8] M. Hauswirth and M. Jazayeri. A component and communication model for push systems. In *ESEC/FSE '99*, pages 20–38. Springer-Verlag, 1999.

[9] K. Juvva and R. Rajkumar. A real-time push-pull communications model for distributed real-time and multimedia systems. Technical Report CMU-CS-99-107, School of Computer Science, Carnegie Mellon University, January 1999.

[10] R. Khare. Beyond Ajax: Accelerating web applications with real-time event notification. Knownow.com, white-paper.

[11] R. Khare and R. N. Taylor. Extending the representational state transfer (REST) architectural style for decentralized systems. In *ICSE '04: Proceedings of the 26th International Conference on Software Engineering*, pages 428–437. IEEE Computer Society, 2004.

[12] J.-P. Martin-Flatin. Push vs. pull in web-based network management. http://arxiv.org/pdf/cs/9811027, 1999.

[13] A. Mesbah and A. van Deursen. An architectural style for Ajax. In *WICSA '07: Proceedings of the 6th Working IEEE/IFIP Conference on Software Architecture*, pages 44–53. IEEE Computer Society, 2007.

[14] A. Mesbah and A. van Deursen. Migrating multi-page web applications to single-page Ajax interfaces. In *CSMR '07: Proceedings of the 11th European Conference on Software Maintenance and Reengineering*, pages 181–190. IEEE Computer Society, 2007.

[15] Netscape. An exploration of dynamic documents. http://wp.netscape.com/assist/net_sites/pushpull.html, 1996.

[16] A. Russell. Comet: Low latency data for the browser. http://alex.dojotoolkit.org/?p=545.

[17] A. Russell, G. Wilkins, and D. Davis. Bayeux - a JSON protocol for publish/subscribe event delivery protocol 0.1draft3. http://svn.xantus.org/shortbus/trunk/bayeux/bayeux.html, 2007.

[18] R. Srinivasan, C. Liang, and K. Ramamritham. Maintaining temporal coherency of virtual data warehouses. In *RTSS '98: Proceedings of the IEEE Real-Time Systems Symposium*, page 60. IEEE Computer Society, 1998.

[19] V. Trecordi and G. Verticale. An architecture for effective push/pull web surfing. In *2000 IEEE International Conference on Communications*, volume 2, pages 1159–1163, 2000.

[20] M. Welsh, D. Culler, and E. Brewer. Seda: an architecture for well-conditioned, scalable internet services. *SIGOPS Oper. Syst. Rev.*, 35(5):230–243, 2001.

[21] M. Welsh and D. E. Culler. Adaptive overload control for busy internet servers. In *USENIX Symposium on Internet Technologies and Systems*, 2003.

---

[15] http://www.lightstreamer.com

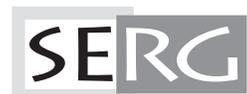